\documentclass[manuscript]{emulateapj}

\slugcomment{to appear in the Astrophysical Journal}

\shorttitle{Light-Curve Model of PU Vul}
\shortauthors{Kato et al.}

\begin{document}


\title{A Light-Curve Model of the Symbiotic Nova PU Vul (1979)
 -- A Very Quiet Explosion with Long-Lasted Flat Peak}


\author{Mariko Kato}
\affil{Department of Astronomy, Keio University, Hiyoshi, Yokohama
  223-8521 Japan}
\email{mariko@educ.cc.keio.ac.jp}

\author{Izumi Hachisu}
\affil{Department of Earth Science and Astronomy, College of Arts and
Sciences, University of Tokyo, Komaba, Meguro-ku, Tokyo 153-8902, Japan}

\author{Angelo Cassatella}
\affil{INAF, Istituto di Fisica dello Spazio Interplanetario,
 Via del Fosso del Cavaliere 100, 00133 Roma, Italy}
\affil{European Space Astronomy Center (ESAC), Villanueva de la Can\~ada,
Apartado 78, 28691, Madrid, Spain}
\affil{Dipartimento di Fisica E. Amaldi, Universit\`a degli Studi Roma Tre,
Via della Vasca Navale 84, 00146 Roma, Italy}

\and

\author{Rosario Gonz\'alez-Riestra} \affil{ { XMM Sciece Operation
 Centre, ESAC, P.O. Box 78, 28091 Vallanueva de la Ca\~nada, Madrid,
 Spain} } 




\begin{abstract}
We present a light curve model 
of the symbiotic nova PU Vul (Nova Vulpeculae 1979) that shows 
a long-lasted flat peak with no spectral indication of 
wind mass-loss before decline. 
Our quasi-evolution models consisting of a series of static solutions  
explain both the optical flat peak and 
ultraviolet (UV) light curve simultaneously. 
The white dwarf mass is estimated to be $\sim 0.6 M_\odot$.  
We also provide a new determination of the reddening, $E(B-V) = 0.43 \pm 0.05$,  
from UV spectral analysis. 
Theoretical light curve fitting of UV 1455 \AA~ provides the 
distance of $d=3.8 \pm 0.7$ kpc.
\end{abstract}




\section{Introduction} \label{sec_introduction}

PU Vul was independently discovered by Y. Kuwano \citep{koz79a} and 
M. Honda \citep{koz79b} as a nova-like object. 
Subsequent observations revealed that PU Vul did not behave like 
a classical nova in its photometric and spectroscopic properties.  
After the rise to the optical maximum, PU Vul maintained 
an almost stable maximum of $V=8.6$ from 1979 to 1987 except for a deep 
minimum in 1980. 
It started to fade in 1988 very slowly toward the prediscovery magnitude 
of $B=14.5-16.6$ \citep{lil79} and the photographic magnitude 
$m_{\rm pg}=15.0-16.5$ \citep{yam82}. 
Such a long-lasted flat peak and a very slow evolution of the light curve 
made this object quite different from ordinary novae.
PU Vul was recognized to be a binary system consisting 
of an M giant and an outbursting component \citep{bel82a,fri84} and also 
that the outburst is a thermonuclear runaway event on a white dwarf (WD)  
in a symbiotic binary \citep{ken86}.
There were debates on the origin of the deep minimum occurred in 1980  
\citep[see discussion in][]{ken86}. After the second eclipse in 1994, 
it has been clear that PU Vul is an eclipsing binary with the orbital period of 13.4 yr. 
The observational properties are summarized in Table \ref{table_observation}, 
that shows the discovery date, nova speed class, 
remarkable property of the light curve, indication of dust 
formation, observational evidence of eclipse, orbital period, 
peak magnitudes of $m_V$, 
information of UV 1455 \AA~ continuum-band light-curve obtained from fitting 
of Model 1 (Section \ref{sec_UVlightcurve}), i.e., peak value and full width 
of half maximum (FWHM), extinction in literature and our estimates, distance to 
PU Vul in literature and our estimates.


\begin{deluxetable*}{lllll}
\tabletypesize{\scriptsize}
\tablecaption{Observational Properties of PU Vul
\label{table_observation}}
\tablewidth{0pt}
\tablehead{
\colhead{subject} &
\colhead{} &
\colhead{data} &
\colhead{units} &
\colhead{comments}
}
\startdata
discovery date& ... &  5.82 April 1979 & UT &  \citet{koz79a} \\
nova speed class & ... & very slow & &    \\
light curve & ... & flat peak &  & \\
dust &... & no && \\
eclipse & ... & yes & &  \\
orbital period & ... & 13.4 & yr &  \\
$m_{V}$ of flat peak  &... & 8.6  & mag & \\
peak of UV 1455 \AA~flux &... & $6.0 \times 10^{-13}$& erg~cm$^{-2}$~s$^{-1}$\AA$^{-1}$ & this work: Section \ref{sec_lightcurvefitting} \\ 
FWHM of UV 1455 \AA &... & 4.4 & yr & this work: Section \ref{sec_lightcurvefitting}\\ 
$E(B-V)$ & ... &0.29-0.5 & &  see references\tablenotemark{a}  \\
$E(B-V)$ & ... &$0.43 \pm 0.05$ & &  this work: Section \ref{sec_UVreddening}\\ 
distance &... & 1.6-7 & kpc &  see references \tablenotemark{b}\\
distance &... & $3.8 \pm 0.7 $ & kpc &  this work: Section \ref{sec_distance}
\enddata
\tablenotetext{a}{$E(B-V)=0.4$ \citep{bel82b}, 0.49 \citep{fri84}, 0.4--0.5 \citep{ken86},
 0.50 \citep{goc91}, 0.4 \citep{vog92}, 0.29 \citep{lun05}}
\tablenotetext{b}{$d=5$--7 kpc \citep{bel82b}, 5.3 kpc \citep{bel84}, $<$ 5.6 kpc
  \citep{goc91}, 1.6--2.0 kpc \citep{vog92}, 2.5 kpc \citep{hoa96}}
\end{deluxetable*}

Spectral development of PU Vul was extensively studied by various 
authors \citep{yam82,yam83,iij84,bel89,goc91,kan91a,kan91b,tam92,kle94}. 
The spectra mimicked those of an F supergiant 
 \citep{yam82,kan91a} in the early phase and changed to A0 \citep{bel89,vog92} from 1983 to 1986 
as the excitation temperature gradually increased \citep{kan91b}.
\citet{yam82} commented that the eruption must have been quite soft because 
no evidence of shell ejection, both in emission lines and shell 
absorption lines, was detected. 
The optical spectrum  was strongly absorption-dominated until 1985 but 
changed to a distinct nebular spectrum in the second half of 1987 
\citep{iij89,kan91b}.
In 1990 the star had shown rich emission lines in the optical and UV spectrum,   
which are typical in the nebular phase and associated with an 
extended atmosphere of a WD \citep{vog92,kan91b,tom91}.

It is very interesting that there is no indication of strong winds in PU Vul 
in contrast to many other classical novae. 
Instead, optically thin mass-ejection from WD photosphere was suggested from 
P Cygni line-profiles \citep{bel89,vog92,sio93b,nus96} 
or triple structure of IR emission lines \citep{ben91}.
The line width corresponds to 1100--1200 km~s$^{-1}$ in average full widths at zero 
intensity \citep{tom91}, $\sim 2600$ km~s$^{-1}$ in Balmer emission wings \citep{iij89},
and 550--600 km~s$^{-1}$ in UV spectra \citep{sio93b}. 
Mass ejection was also suggested from X-ray emission 
detected with ROSAT on 10--12 November 1992 UT and 
interpreted as thermal bremsstrahlung  with a temperature of 
0.22 keV ($2.6 \times 10^6$ K) \citep{hoa96,mue97}. These authors suggested a  
collisional origin of the X-ray between a high-density, low-velocity cool wind
from the M giant and a low-density, high-velocity hot wind from the 
WD in the context of common properties of symbiotic novae.

To summarize, the outburst of PU Vul was very quiet in the first ten years,  
and optically thin mass-ejection arises from 1988--1990 as the 
nova entered a coronal phase.  
These spectral features as well as the long-lasted flat peak 
make PU Vul quite different from many other classical novae 
in which optical magnitude decays quickly from its peak and spectrum 
indicates strong optically thick winds. 
This paper aims to model such a quite different evolution of PU Vul 
and to understand the cause of such properties.

\citet{kat09} reexamined the conditions of occurrence of optically thick winds 
and found that optically thick winds occur in a limited range of 
the envelope (ignition) mass. For a relatively large envelope mass, 
optically thick winds are suppressed 
in a way that a large density-inversion layer appears and the gas-pressure 
gradient balances with the radiation-pressure gradient, the driving force of the winds. 
In massive WDs ($\gtrsim 0.7~M_\odot$), optically thick winds always 
occur, because the ignition mass of the 
wind-suppressed solutions are too massive to be 
realized in the actual novae.
In less massive WDs ($\lesssim 0.5~M_\odot$), on the other hand, 
no winds are accelerated because the radiation-pressure gradient is 
too weak to drive the winds. 
In between them, i.e., $0.5~M_\odot \lesssim M_{\rm WD} \lesssim 0.7~M_\odot$, 
both types of solutions (wind and wind-suppressed) can be realized depending 
on the initial envelope mass. For a less massive envelope,  
optically thick winds are accelerated and a shell flash develops 
as a normal nova with strong winds. On the other hand,  
if the initial envelope mass is relatively large, optically thick winds 
are suppressed and a nova evolves without winds. 

\citet{kat09} also presented an idea that such wind-suppressed (no optically 
thick wind) evolution will be realized in a very slow nova that shows 
a long-lasted flat optical peak. 
If the optically thick wind occurs, as in many classical novae, 
the strong winds carry out most of the envelope matter in a short timescale, 
and the optical brightness quickly decays. Thus, the light 
curve has a sharp optical peak.  
On the other hand, in the wind-suppressed evolutions, 
the brightness decays very slowly, because the evolution timescale
is determined only by hydrogen nuclear burning. Therefore, the nova stays  
at a low surface temperature for a long time, which makes 
a long-lasted flat optical peak. 
PU Vul is the first example of this new type of evolution.

In Section \ref{sec_UV} we  review observational results based on the 
{\it IUE} spectra.  
In Section \ref{sec_model} we briefly introduce our method and assumptions of  
the theoretical model. 
Section \ref{sec_lightcurvefitting} shows how to estimate the WD mass from 
light curve fittings. 
Discussion and conclusions follow
in Sections \ref{sec_discussion} and \ref{sec_conclusions}, respectively.

\section{UV Observations } \label{sec_UV}

PU Vul had been monitored by {\it IUE} from February 1979 to September
1983  and from October 1987 to September 1996 at both low and high
resolutions. A gallery of UV SWP spectra from 1992 to 1995 can be found
in \citet{nus96}.

In the following we revisit the problem of the color excess $E(B-V)$ of PU Vul, 
and describe the long term evolution of the emission lines. The evolution of 
the UV continuum will be described in Section \ref{sec_UVlightcurve}. 
The ultraviolet spectra were retrieved from the {\it IUE}
archive through the INES ({\it IUE} Newly Extracted Spectra)
system\footnote{http://sdc.laeff.inta.es/ines/}, which also provides full
details of the observations.  The use of {\it IUE} INES data is particularly
important for the determination of reddening correction because of the
implementation of upgraded spectral extraction and flux calibration procedures
compared to previously published UV spectra.

\subsection{Reddening Correction} \label{sec_UVreddening}

Table \ref{table_observation} shows that the color excess $E(B-V)$ 
toward PU Vul in the literature lies in the range from 0.3
to 0.5. Given the large spread of these determinations, we have directly 
determined  $E(B-V)$ from the strength of the 2200 \AA~ feature seen in the 
UV spectra of PU Vul.

\begin{figure}
\epsscale{1.15}
\plotone{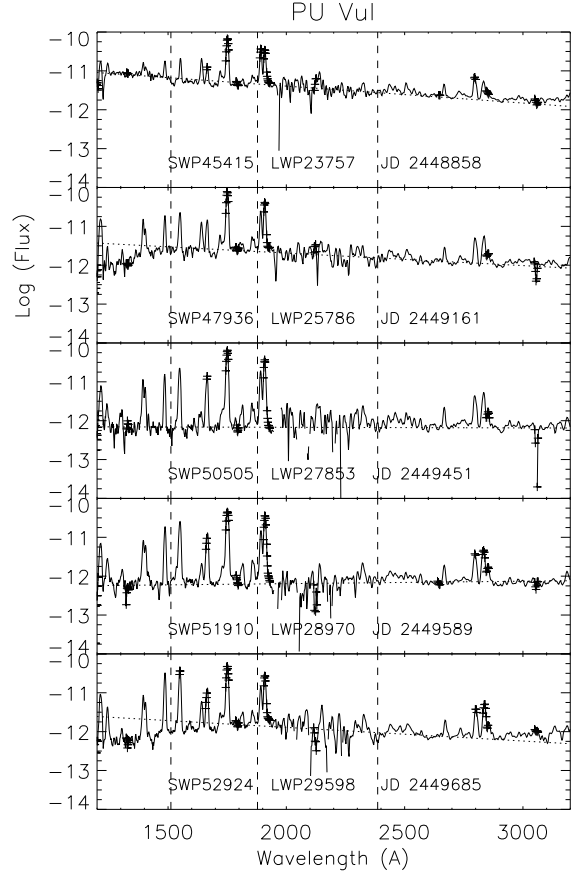}
\caption{{\it IUE} spectra of PU Vul obtained at different dates.
Fluxes are in units of erg cm$^{-2}$ s$^{-1}$ \AA$^{-1}$. The spectra
have been corrected for reddening using $E(B-V)=0.43$.  The vertical
dotted lines represent the wavelengths $\lambda\lambda$ 1512, 1878 and
2386 \AA~ at which the extinction law takes the same value.  With the
adopted value of reddening, the stellar continuum underlying the many
emission lines is well represented by a straight line all over the
full spectral range.  Saturated data points in the emission lines are
labeled with pluses.
}
\label{fig:scattering}
\end{figure}

The Galactic extinction curve \citep{sea79} shows a pronounced broad maximum
around 2175 \AA~ due to dust absorption.  Since  it takes the same value
$X(\lambda) = A(\lambda)/E(B-V) \approx 8$ at $\lambda = 1512$, 1878, and 2386
\AA,  the slope of the straight line passing through the continuum
points at these wavelengths is insensitive to $E(B-V)$ in a $(\lambda , \log
F(\lambda))$ plot.  This circumstance can be used to get a reliable estimate
of $E(B-V)$ as that in which the stellar continuum becomes closely linear in
the 1512--2386 \AA~ region, and passes through the continuum points at the
above wavelengths. From 9 pairs of short and long wavelength {\it IUE}
spectra taken from JD 2,448,217 to JD 2,450,342, i.e. during
the nebular phase, we have in this way found $E(B-V) = 0.43 \pm 0.05$. 
Examples of {\it IUE} spectra of PU Vul corrected with $E(B-V) =
0.43$ are reported in Figure \ref{fig:scattering}.


\begin{figure}
\epsscale{1.15}
\plotone{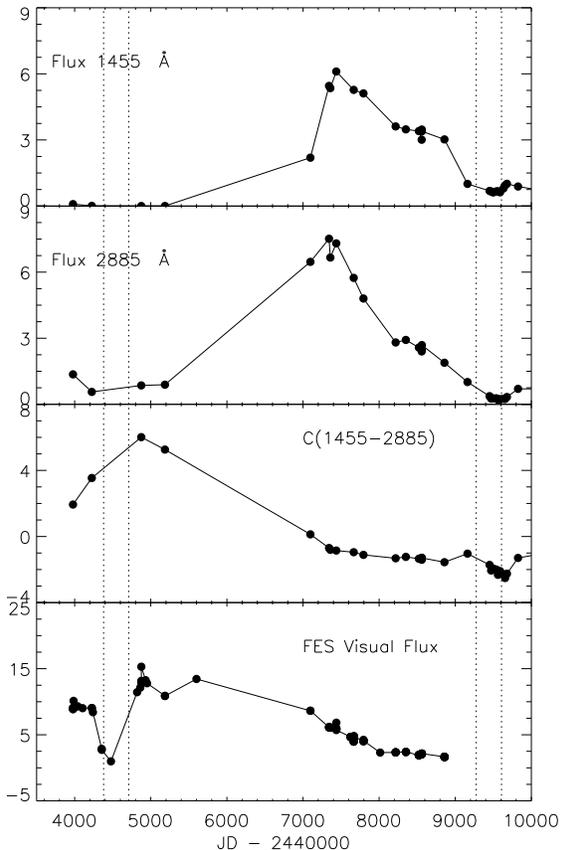}
\caption{Evolution of the continuum fluxes at 1455 \AA\ and 2885 \AA,
  of the ultraviolet color index $C(1455 - 2885)$, and of the
  $V_{\rm FES}$ visual flux of PU Vul (see Section \ref{uv_evolution_cont}). 
  Visual and UV fluxes are both in
  units of 10$^{-13}$ erg cm$^{-2}$ s$^{-1}$ \AA$^{-1}$, not corrected
  for reddening.  Only the color index has been corrected for
  reddening using $E(B-V) = 0.43$ (see Section \ref{sec_UVreddening}). 
  The vertical dotted lines indicate the period of the first 
  (JD 2,444,380 -- 2,444,710) and the second 
  (JD 2,449,275 -- 2,449,608) eclipses.  }
\label{fig:cont_puvul}
\end{figure}

\subsection{Evolution of the UV Continuum}
\label{uv_evolution_cont}

We have measured the mean flux in two narrow bands 20 \AA\ wide
centered at 1455 \AA\ and 2855 \AA, selected to provide a fairly
good representation of the UV continuum because little affected by emission
lines \citep{cas02}. Figure \ref{fig:cont_puvul} shows the time
evolution of the $F(1455$~\AA) and $F(2885$ \AA) fluxes and of the UV
color index $C(1455-2885) = -2.5 \log[F(1455$~\AA)$/F(2885$~\AA)].
The measurements were made on well exposed low resolution large
aperture spectra.  Figure \ref{fig:cont_puvul} reports also, for
comparison, the visual light curve obtained from the Fine Error Sensor
(FES) counts, $V_{\rm FES}$, on board {\it IUE}, 
once corrected for the time dependent
sensitivity degradation \citep[p.35 in][]{cas04}.
Note the pronounced UV maximum around JD
2,447,400 followed by a progressive decay at the same time as the UV
spectrum becomes harder, as indicated by the decrease of the UV color
index. The delay of the UV maximum with respect to the visual maximum
is common both to novae and symbiotic stars \citep{fer95,cas02}.


\begin{figure}
\epsscale{1.15}
\plotone{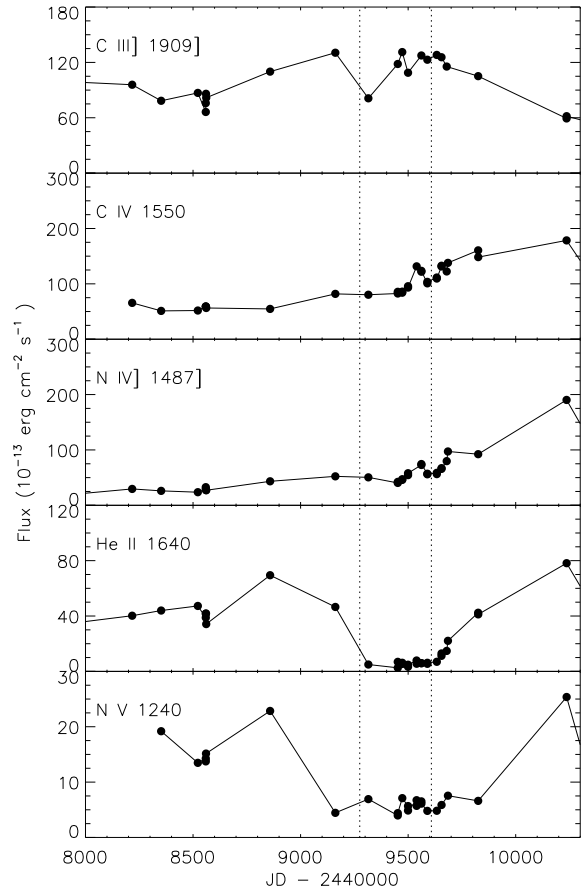}
\caption{Observed fluxes of some prominent UV emission lines of PU Vul as
  a function of time.  Fluxes are in units of 10$^{-13}$ erg cm$^{-2}$ s$^{-1}$,
  not corrected for reddening.
}
\label{fig:lines_puvul}
\end{figure}

\subsection{Evolution of the UV Emission Lines}

Figure \ref{fig:lines_puvul} reports, as a function of time, the
observed fluxes in the emission lines of C III] 1909 \AA, C IV 1550 \AA, N
IV] 1487 \AA, He II 1640 \AA, and N V 1240 \AA, which cover a wide
range of ionization conditions (the corresponding ionization energies
${\chi_{ion}}$ are 24.4, 47.9, 47.4, 54.4 and 77.5 eV,
respectively).  The flux measurements were obtained from well exposed
{\it IUE} low resolution spectra. The figure shows clearly that the
eclipse around JD 2,449,550 endures longer for high than for low ionization
lines.  This suggests that the high ionization lines are formed in
the unseen side of the cool giant's wind, so confirming the results
by \citet{nus96} (see their Figure 5), who found that the UV highest
excitation lines of He II $\lambda 1640$, N V $\lambda 1240$, and N IV
$\lambda 1718$ disappeared during the second eclipse.

\section{Model}
\label{sec_model}

\subsection{Evolution of Nova Outburst} \label{subsec_model}

Nova is a thermonuclear runaway event on a WD \citep{nar80,pri86,sta72,sta88,jos98,pri95}. 
After unstable hydrogen nuclear burning triggers a nova outburst, 
the envelope on the WD greatly expands to a giant size. 
After it reaches the optical peak, the envelope expansion settles down into a steady-state. 
The optical magnitude decreases as the envelope mass decreases and 
the photospheric temperature rises with time. 
In less massive WDs the optically thick wind 
does not occur as described in Section \ref{sec_introduction}
and, therefore, the decay phase of novae can be followed by a
quasi-hydrostatic sequence \citep[e.g.,][]{ibe82,kat94h,kat09}. 
We solved the equations of hydrostatic balance, radiative diffusion, and 
conservation of energy from the bottom of the hydrogen-rich 
envelope through the photosphere.  
The bottom radius is assumed to be the Chandrasekhar radius. 
The evolution of novae is followed by connecting these solutions 
along the envelope mass-decreasing sequence. 
The time evolution is calculated from the 
mass decreasing rate which is the summation of the two  
rates, hydrogen nuclear burning  and optically-thin wind mass-loss. 
We used  OPAL opacities \citep{igl96}.
The method and numerical techniques are essentially the same as those  
in \citet{kat94h}. 
Convective energy transport is calculated using the mixing length theory with 
the mixing-length parameter $\alpha=1.5$ 
\citep[see Figure 11 in][for the dependence of model light curve on the mixing
length parameter $\alpha$]{kat09}. 

In the rising phase, the envelope does not yet settle down to a thermal equilibrium,
i.e., the nuclear energy generation is larger than the radiative loss.
We have approximated such a stage by a sequence of static solutions of 
constant mass without thermal equilibrium. 
These solutions may not approximate well the rising phase, 
but are enough to our purpose, because the rising phase plays no important role 
in the determination of physical values such 
as the WD mass and distance.

\subsection{Wind Mass-Loss Rate}

In the later phase of the outburst (after 1986), 
optically-thin wind may arise because many emission lines had appeared in spectra. 
We cannot calculate such wind mass-loss, however,  
because radiative transfer in the optically-thin region is not included 
in our model calculation. Therefore, we take the wind mass-loss rate 
as a model parameter. This optically-thin wind does not affect the envelope 
structure below the photosphere, but speeds up the nova 
evolution because mass decreasing rate of 
the envelope is accelerated by this wind mass-loss. 

In line driven winds, the mass-loss rate $\dot M$ is limited by 
photon momentum; the wind cannot get momentum much exceeding the momentum
of photon flux. This is in contrast to optically thick winds in which 
the mass-loss rate could be much larger than the momentum of photon flux \citep{kat92}. 
Therefore,  we assume a condition, that the momentum of wind is smaller 
than that of photon flux, of   

\begin{equation}
\dot M v_{\rm wind} < {L_{\rm ph} \over c},
\label{equation_masslossC}
\end{equation}
where $c$ is the speed of light.   We get an upper 
limit of the wind mass-loss rate as 

\begin{equation}
\dot M  <  1.6 \times 10^{-6}~
({L_{\rm ph} \over 6 \times 10^{37}{\rm erg~s}^{-1}})
({v_{\rm wind}\over 200 {\rm~km~s}^{-1}})^{-1} ~M_\odot~{\rm yr}^{-1}.
\label{equation_masslossrate}
\end{equation}
Here we take $L_{\rm ph} =6 \times  10^{37}$ erg~s$^{-1}$ from a bolometric 
luminosity of a 0.6 $M_\odot$ WD (which will be shown later as Model 2  
in Table \ref{table_model})
, and assume a relatively small value of 
$v_{\rm wind}=200$ km~s$^{-1}$ for a safe upper limit.

\citet{ben91} estimated the mass-loss rate of PU Vul 
to be about $4 \times 10^{-6}~M_\odot$~yr$^{-1}$
from emission measure on 30 April and 1 May 1988 with a wind velocity of 
70 km~s$^{-1}$ and a  hot star radius of $3.9 \times 10^{12}$ cm ($56~R_\odot$).  
\citet{sio93b} set an upper limit of the mass-loss rate, $1 \times 
10^{-5}~M_\odot$~yr$^{-1}$. \citet{sko06} estimated the mass-loss rate
from $H\alpha$ line luminosity of 144 $L_\odot$ on 28 September 1988, assuming 
optically-thin fully-ionized winds.  
With a terminal velocity 2,100 km~s$^{-1}$ of a given velocity profile, 
Skopal obtained the mass-loss rate to be 
$\dot M = 5\times 10^{-6}M_\odot~$yr$^{-1}$. 
These values, however, may be a little bit larger than 
our momentum condition (\ref{equation_masslossC}). 

In our model, we assume no wind mass-loss during the optical 
flat peak because no emission lines are observed or they are very weak. 
We assume that optically thin-wind begins when the photospheric temperature 
rises to $\log T_{\rm ph}$ (K) $\sim 4.0$. 
This  wind seems to have much weakened sometime around 1999-2001, 
because, after that, the optical light curve changed its shape and 
seems to decay as $t^{-3}$, where $t$ is the time after the decay started.  
In classical novae, the $t^{-3}$ decay is often observed in the later phase  
of the outburst, and is interpreted as emission from homologously 
expanding optically thin plasma with a constant mass, 
which indicates that the mass supply had stopped \citep{hac06a,hac10}. 
In the case of PU Vul, the wind had not stopped entirely since P Cyg profiles are 
still observed in 2004 \citep{yoo07}.
Therefore, we assume that the optically-thin wind begins at  $\log T_{\rm ph}$ (K) $= 4.0$
and continues until $\log T_{\rm ph}$ (K) $= 5.05$ at a rate shown later, and 
after that the wind mass-loss rate dropped to $\dot M = 1.0 \times 10^{-7} M_\odot$ yr$^{-1}$.

\subsection{Multiwavelength Light Curves}

After the
maximum expansion of the photosphere, the photospheric radius ($R_{\rm ph}$)
gradually decreases keeping the total luminosity ($L_{\rm ph}$)
almost constant.  The
photospheric temperature ($T_{\rm ph}$) increases with time because of
$L_{\rm ph} = 4 \pi R_{\rm ph}^2 \sigma T_{\rm ph}^4$.  The maximum
emission shifts from optical to supersoft X-ray through ultraviolet
(UV).  This causes the luminosity decrease in optical and increase in UV
and finally increase of supersoft X-ray.
We assume that photons are emitted at the photosphere as a
blackbody with a photospheric temperature  $T_{\rm ph}$.  The light curve
of optical ($V$) and UV 1455\AA~ fluxes are estimated from
the blackbody emission.

\subsection{Chemical Composition}
\label{sec_composition}

\citet{bel89} obtained chemical composition of the atmosphere of the hot
component of PU Vul, in which iron is depleted by a factor of 0.3--0.5 against 
the sun. 
The number ratio of He/H is estimated to be 0.31 \citep{and94} and 0.146
\citep{lun05} from emission line ratios, which shows helium overabundance
than the solar value (He/H $\sim 0.08$).
Another suggestion comes from the location in the Galaxy.
\citet{bel82b} suggested that PU Vul does not belong to the planar component of the 
Galaxy because the star is off the galactic plane by 0.7--1.0 kpc, based on their
derived distance of 5--7 kpc.
This value, however, reduced to 0.5 kpc if we adopt the distance of
$\sim 3.8$ kpc as we will obtain later.

With the information above we assume the mass fraction of hydrogen, helium  and heavy
elements of the envelope to be ($X$, $Y$, $Z$)=(0.5, 0.494, 0.006).
For comparison, we further assume additional sets of composition i.e.,
(0.5, 0.49, 0.01), (0.7, 0.28, 0.02) and (0.7, 0.29, 0.01).
We simply assumed that the chemical composition of the envelope is uniform 
and constant with time.

\section{Light Curve Fitting} \label{sec_lightcurvefitting}

\subsection{UV Light Curve Fitting and the WD mass} \label{sec_UVlightcurve}

 The UV 1455 \AA~ flux provides a good representation of
the continuum level in novae, because it coincides with a local minimum of 
line opacity \citep{cas02}.  In previous papers
\citep{hac06a,hac08,kat09v838her}, we have shown that the UV
  1455 \AA~ continuum light curve is very sensitive to model parameters, 
especially the WD mass.


\begin{figure}
\epsscale{1.15}
\plotone{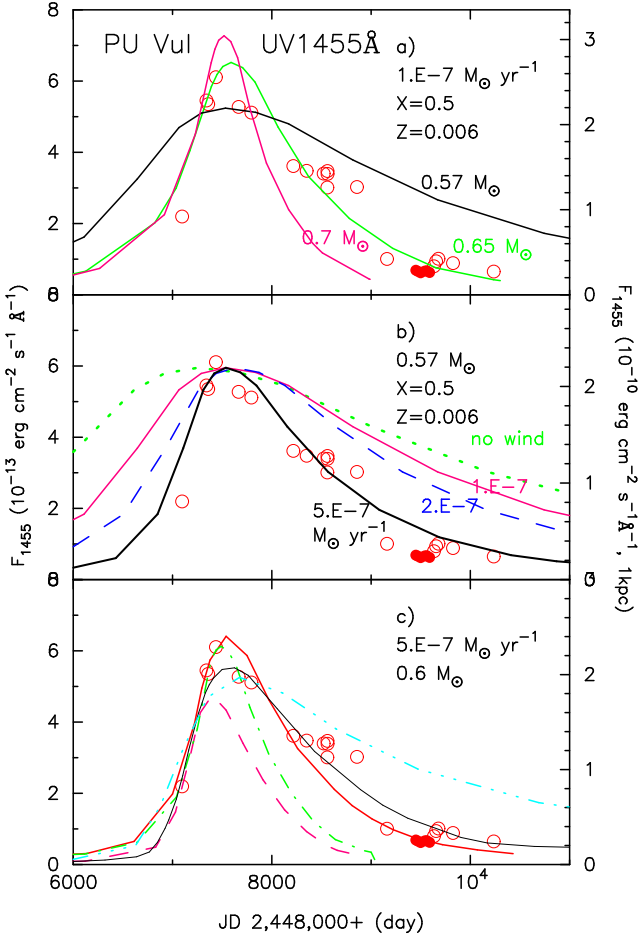}
\caption{UV 1455 \AA~light curves. The 
UV 1455 \AA~ data are denoted by red open circles (outside the eclipse) and 
red filled circles (during the eclipse). 
Theoretical UV light curves are shown in curves for an assumed distance of 1 kpc 
(with right-side axis). 
a) Comparison among different WD masses for the same wind mass-loss rate ($ 1 \times 10^{-7}
M_\odot$ yr$^{-1}$) and the same chemical composition of $X=0.5$ and $Z=0.006$.
 The WD mass is indicated beside each line.
b) Comparison among different mass-loss rates 
for the 0.57 $M_\odot$ WD with composition 
of $X=0.5$ and $Z=0.006$. The wind mass-loss rate 
 (in units of $M_\odot$ yr$^{-1}$) is attached to each curve. 
c) Comparison among different chemical compositions for the same mass-loss rate 
of  $5 \times 10^{-7}M_\odot$ yr$^{-1}$ and WD mass of 0.6 $M_\odot$.
The chemical composition is ($X,Z$)=
(0.7,0.004) for a three-dot-dashed curve, 
(0.5,0.006) for a thick solid curve, 
(0.7,0.01) for a thin solid curve, 
(0.5,0.01) for a dash-dotted curve, and 
(0.7,0.02) for a dashed curve. 
\label{UVwidth}}
\end{figure}

Figure \ref{UVwidth} shows time evolution of the UV 1455 \AA~  continuum flux as  
well as our theoretical light curves. 
Figure \ref{UVwidth}a shows dependence on the WD mass 
for a given set of the wind mass-loss rate of $ 1 \times 10^{-7}M_\odot$ 
yr$^{-1}$ (arbitrarily chosen but not too small compared with the nuclear burning 
rate of $3 \times 10^{-7}M_\odot$ yr$^{-1}$) and chemical composition of 
$X=0.5$ and $Z=0.006$. 
We see that the UV peak is narrower in more massive WDs, because nova evolves 
faster owing to a less massive envelope and a high nuclear burning rate. 
Figure \ref{UVwidth}b depicts four light curves with different wind 
mass-loss rates  for a given 
WD mass and chemical composition. As the assumed wind mass-loss rate is comparable to 
the nuclear burning rate ($3 \times 10^{-7}M_\odot$ yr$^{-1}$), 
the evolution speed of nova is sensitive to the mass-loss rate.
Figure \ref{UVwidth}c shows five light curves with different sets of 
chemical composition. Nova evolves faster for smaller $X$, because of less 
nuclear fuel,  and also faster for larger $Z$ because of a smaller envelope mass.


\begin{deluxetable*}{llllll}
\tabletypesize{\scriptsize}
\tablecaption{Summary of Our Models
\label{table_model}}
\tablewidth{0pt}
\tablehead{
\colhead{subject} &
\colhead{} &
\colhead{model 1} &
\colhead{model 2} &
\colhead{} &
\colhead{units}
}
\startdata
$X$ & ... & 0.5  & 0.5&   & \\
$Y$ & ... & 0.494  & 0.494&  &\\
$Z$ & ... & 0.006  & 0.006&    &\\
WD mass & ... & 0.57&  0.6  &&    $M_\odot$ \\
distance from UV fit \tablenotemark{a}&...& $ 3.7$ & $3.9 $&  & kpc \\
$M_{V, {\rm peak}}$\tablenotemark{b}  & ... &-5.49 &-5.58 & &mag \\
$L_{\rm peak}$\tablenotemark{b} &... &  5.2   &5.7 &  &   $10^{37}$erg~s$^{-1}$ \\
H-burning rate \tablenotemark{c}& ... &2.7  &3.0 && $10^{-7} M_\odot$yr$^{-1}$ \\
maximum radius & ... & 62& 63  &   & $R_\odot$ \\
initial envelope mass \tablenotemark{d}& ... & 5.8 &4.6 &  & $10^{-5} M_\odot$  \\
assumed wind mass-loss rate ($T<5$)\tablenotemark{e}& ...& 5.0 &3.0& $$ & $10^{-7}~M_\odot$~yr$^{-1}$\\
assumed wind mass-loss rate ($T>5$)\tablenotemark{f}& ...& 1.0 &1.0& $$ & $10^{-7}~M_\odot$~yr$^{-1}$\\
mass lost by the wind  ($T<5$)\tablenotemark{e}& ... & 0.60 &0.34  &  & $10^{-5}M_\odot$ \\
mass lost by the wind  ($T>5$)\tablenotemark{f}& ... & 0.61 &0.44  &  & $10^{-5}M_\odot$ 
\enddata
\tablenotetext{a}{with $E(B-V)=0.43$}
\tablenotetext{b}{values at $\log T_{\rm ph}$(K)=3.9}
\tablenotetext{c}{values at $\log T_{\rm ph}$(K)=4.5}
\tablenotetext{d}{the mass at the rising phase.}
\tablenotetext{e}{optically-thin wind from $\log T_{\rm ph} $(K) = 4 to 5.05.} 
\tablenotetext{f}{optically-thin wind at $\log T_{\rm ph} $(K) $> 5.05$}.  
\end{deluxetable*}

In this way we can choose a WD mass with reasonable agreement with
the UV data for a given parameter set of the wind mass-loss rate and composition, 
($X$, $Y$, $Z$)=(0.5, 0.494, 0.006). 
Table \ref{table_model} shows two such models, one is a 0.57 $M_\odot$ WD
with the wind mass-loss rate of $5 \times 10^{-7}M_\odot$~yr$^{-1}$ (Model 1) and the
other is a 0.6  $M_\odot$ WD with $3 \times 10^{-7}M_\odot$~yr$^{-1}$ (Model 2).
These WD masses, 0.57 $M_\odot$ and 0.6  $M_\odot$, are not much different, 
because they are in the middle of the permitted range of the WD mass, 0.53--0.65 
$M_\odot$, as explained below to avoid extremely small and large mass-loss rates.
The 0.57 and 0.6 $M_\odot$ models correspond respectively to the left and upper sides  
to the ``wind region'' \citep[a triangle region in Figure 10 of][]{kat09} 
for the corresponding composition.


\begin{figure}
\epsscale{1.15}
\plotone{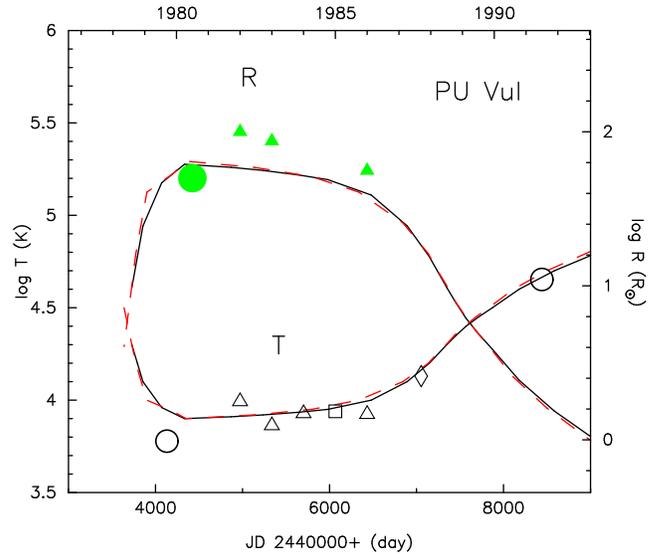}
\caption{Development of the temperature and radius of the hot component
(WD photosphere)
in PU Vul. Curves denote the blackbody temperature and the photospheric
radius of our Model 1 (Solid line) and  Model 2 (Dashed line). 
The effective temperature (open symbols) and radius (filled symbols) 
estimated by various authors are also shown.
Circles: \citet{vog92}. Triangles: \citet{bel89}.
Diamonds: \citet{goc91}. Square: \citet{ken86}.
\label{Temp}}
\end{figure}

During the outburst, the photospheric temperature gradually rises and the
photospheric radius decreases, with an almost constant photospheric luminosity.
Figure \ref{Temp} shows the developments of the temperature and radius of Model
1 and Model 2  as well as observational estimates taken from literature.
\citet{vog92} suggested that the temperature was as low as  6,000 K in 1979
as an effective temperature of F-supergiant
and  estimated the temperature to be 40,000 K $< T <$ 50,000 K at 1991 from
nebular spectra.
They also obtained the radius of $ \sim 50 R_\odot$ from the analysis of 1980
eclipse.
\citet{ken86}, \citet{bel89}, and \citet{goc91} estimated the temperature from spectral type.
These values are plotted in Figure \ref{Temp}, in which
our model temperature well reproduces observational estimates including
the epoch around the UV1455 \AA~ light curve peak, i.e., JD 2,447,000 to 2,448,200.
The model radius is also consistent with observations
considering large ambiguity in the estimating methods.


\begin{deluxetable*}{lllll}
\tabletypesize{\scriptsize}
\tablecaption{Upper and Lower Values of the WD Mass
\label{table_WDmass}}
\tablewidth{0pt}
\tablehead{
\colhead{composition} &
\colhead{} &
\colhead{minimum mass\tablenotemark{a}} &
\colhead{maximum mass\tablenotemark{b}} &
\colhead{}
}
\startdata
$X=0.5,~Y=0.294,~Z=0.006$&...& 0.53 $M_\odot$ &0.65~$M_\odot$ & \\
$X=0.5,~Y=0.29,~Z=0.01$&...& 0.5 $M_\odot$ & 0.62 $M_\odot$ & \\
$X=0.7,~Y=0.28,~Z=0.02$&...& $0.5~M_\odot$ &0.67~$M_\odot$ & \\
$X=0.7,~Y=0.29,~Z=0.01$&...& $0.52~M_\odot$ &0.72~$M_\odot$ & \\
\enddata
\tablenotetext{a}{in case of a very large mass-loss rate of $1\times 10^{-6}~M_\odot~$yr$^{-1}$}
\tablenotetext{b}{extreme case of no wind mass-loss}
\end{deluxetable*}

Table \ref{table_WDmass} shows the upper and lower limits of the WD mass
that reproduces reasonable fitting to the UV light curve.
For a given chemical composition, the maximum WD mass is obtained with
no wind mass-loss,
because more massive WDs produce much narrower UV curves. A more massive
WD than the maximum value in Table \ref{table_WDmass} produces
a UV light curve too narrow to fit the observation.
On the other hand, for less massive WDs, we need to assume larger wind
mass-loss rates because of their slower evolutions.
The minimum WD mass may be obtained for a largest wind mass-loss
rate that we adopt $1 \times 10^{-6}M_\odot$~yr$^{-1}$ from Equation
(\ref{equation_masslossrate}).
With plausible values of the optically thin wind-mass-loss rate, which may be 
a few to several $\times 10^{-7} M_\odot$~yr$^{-1}$, 
we may conclude that the WD is about $0.6 M_\odot$.


\begin{figure}
\epsscale{1.15}
\plotone{f6.epsi}
\caption{Optical and UV light curves of PU Vul.
Optical data are taken from IAU Circular (for dip: 3421, 3477, 3494, 3589,
3604, 3610, and 3655),
\citet{wen79}, \citet{yam82}, \citet{bel82b}, \citet{cho81},
\citet{pur82}, \citet{kol83}, \citet{pur83}, \citet{iij84},
\citet{iij89}, \citet{kan91b}, \citet{kle94},
\citet{kol95}, \citet{yoo00}, and AAVSO (after JD 2,452,000).
Large red open circles denote the flux of {\it IUE} UV 1455~\AA~ band.
Calculated light curves are also shown for Model 1 (Thin solid line) and  
Model 2 (Dashed line). 
The scale in the right-hand-side axis denotes that for observational 
data. Scale for the theoretical UV flux is (bottom, top)=(0, 6.25)
for Model 1 and    (0, 7.0) for Model 2 in units of 
 10$^{-6}$ erg s$^{-1}$ cm$^{-2}$ \AA$^{-1}$  for the distance of 10 pc. 
The cross/small dot indicate starting/end point of the assumed optically-thin wind
mass-loss. See text for more details. 
\label{light}}
\end{figure}

\subsection{Optical Light Curve}

Figure \ref{light} shows optical light curves of Model 1 and Model 2, of which  
characteristic values are summarized in Table \ref{table_model}. 
These two models are selected from fitting with the UV light curves,
but also well reproduce the optical light curve in the flat maximum as 
well as the following decline until 1989,
except the first eclipse in 1980 which is not taken into account in our model.

After 1989 the theoretical light curve largely deviates from observed 
optical magnitudes.
In this stage, spectra are emission-line dominated and the continuum is very
weak \citep{yoo07}. These emission lines comes from
optically thin plasma outside the photosphere which is not included
in our model. Therefore, our theoretical models give much lower magnitudes 
than that of observational data.

In our theoretical models, the flat peak corresponds to the era of
low photospheric temperature
(7,000--9,000 K) as shown in Figure \ref{Temp}. The temperature gradually rises
with time and reaches 10,000 K which is indicated by the cross 
in Figure  \ref{light}, where we assume the optically-thin mass-loss begins.
After that the nova entered the coronal phase and
many emission lines appeared \citep{kan91b,nus96}.
Our model temperature is  consistent
with these observational properties.

\subsection{Internal Structure of the Envelope at the Flat Peak}\label{sec_structure}


\begin{figure}
\epsscale{1.15}
\plotone{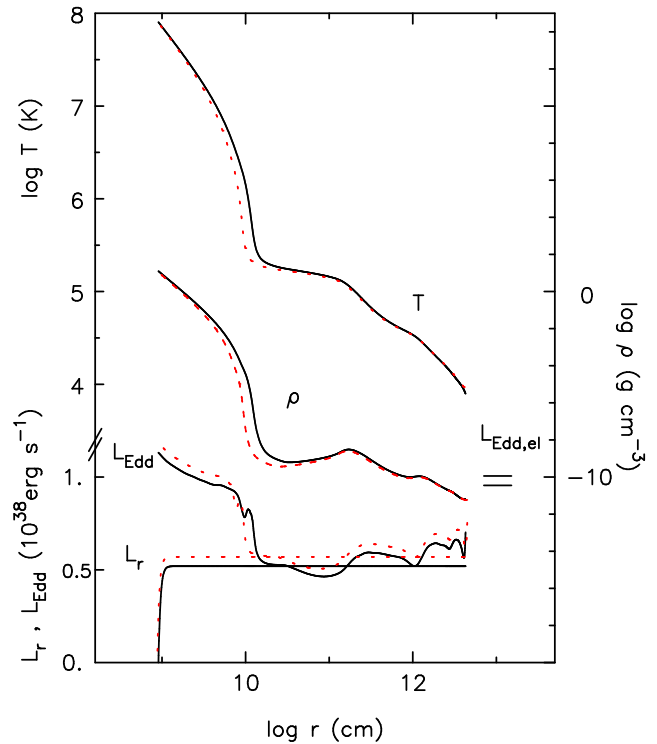}
\caption{Temperature ($T$), density ($\rho$), diffusive luminosity ($L_{\rm r}$), 
and the local Eddington luminosity ($L_{\rm Edd}$)
of envelope solutions at $\log T_{\rm ph}=3.9$ for  Model 1 (solid curve) 
and Model 2 (dotted curve). 
The right edge of the curve corresponds to the photosphere and the left edge to 
the bottom of the envelope, i.e., bottom of the nuclear burning region.
Two short horizontal lines indicate Eddington luminosity with electron scattering 
opacity, $L_{\rm Edd,el}=4 \pi c G M /[0.2(1 + X)]$ for Model 1 (lower line) and Model 2 (upper line).
\label{structure}}
\end{figure}

Figure \ref{structure} shows internal structures of the envelopes of Model 1 
and Model 2 with the photospheric temperature $\log T_{\rm ph}$ (K) $\sim 3.9$.  
Here, the local Eddington luminosity is defined as

\begin{equation}
L_{\rm Edd} \equiv {4\pi cGM_{\rm WD} \over\kappa},
\label{equation_Ledd} \end{equation}
\noindent
where $\kappa$ is the opacity in which we use the OPAL opacity. 
Since the opacity $\kappa$ is a function of temperature and density, 
the Eddington luminosity is also a local variable. 
This Eddington luminosity has the deepest local minimum at $\log R$ (cm) = 10.4 
--11.2 that  corresponds to the Fe peak of OPAL opacity. 
There appears a large density-inversion layer at $\log r$ (cm) $\sim 10.2-11.2$  
corresponding to the super Eddington region ($L_{\rm Edd} < L_r$).
This density-inversion arises in order to keep hydrostatic balance 
in the super-Eddington region.
Such a structure is very different from ordinary nova wind solutions, in which 
the density monotonically decreases as $r^{-2}$ \citep{kat94h,kat09}, but similar 
to that of red giants.

These two solutions are representative of those in the long-lasted flat-peak 
in the optical light curve (Figure \ref{light}), i.e.,  
the envelope is extended to 50--60 $R_\odot$ and the temperature is as low as    
$\log T_{\rm ph}$ (K) $\sim 3.9$--4.0. 
The convection, which is dominant in energy transport in the rising phase
of nova outbursts \citep{pri86}, has retreated and is ineffective 
in and after the flat-peak. 
The convection occurs where the opacity decreases outward, i.e., the local 
Eddington luminosity increases outward.
The largest convective region is at $\log r$ (cm) = 10.5--11.4, corresponding 
to the super-Eddington region (see Figure \ref{structure}). 
In all convective regions, convection is ineffective in energy transport 
due to low density, and unable to carry all of the energy flux.  
Therefore, the structure is super-adiabatic, i.e., entropy decreases outward.
This situation is different from the convective core of intermediate-mass 
main-sequence star or the inner convective envelope of low-mass red giant star, where the
convective energy transport is effective and the temperature gradient is 
very close to the adiabatic gradient \citep{hay62,ibe65}.

The Eddington luminosity with electron scattering opacity $L_{\rm Edd,el} = 
4 \pi c G M/[0.2(1+X)]$ is often used as an easy estimate of the WD luminosity. 
We note, however, that the photospheric luminosity is only 54 \% of  $L_{\rm Edd,el}$
for Model 1 (0.57 $M_\odot$ WD) and 57 \% for 
Model 2 (0.6  $M_\odot$ WD).

\subsection{Distance}
\label{sec_distance}

The distance to PU Vul is obtained from the comparison of the 
1455 \AA~ band flux with the corresponding model fluxes
\citep{hac06a,kat05h, kat07h}.
The flux of Model 1 is $F_{\lambda}^{\rm mod}$ = 2.2 $\times
10^{-6}~(d/10~{\rm pc})^2$~ergs~cm$^{-2}$~s$^{-1}$~\AA$^{-1}$ at the peak. 
The corresponding observed flux is $F_{\lambda}^{\rm obs}$=6.0 $\times
10^{-13}$~ergs~cm$^{-2}$~s$^{-1}$ ~\AA$^{-1}$ as in Figure \ref{light}.
From these values we obtain the distance of $d=3.7$ kpc with  the 
absorption $A_\lambda=8.3\times E(B-V)$ for $\lambda=1455$~\AA~\citep{sea79}, 
here we use $E(B-V)=0.43$ (obtained in Section \ref{sec_UVreddening}).
If we use the upper and lower limit of $E(B-V)=0.43 \pm 0.05$, we get 
the distance of $d= 3.7^{+0.7}_{-0.6}$ kpc.  
The error coming from the UV flux fitting is much smaller and up to $\pm 0.15 $ kpc.  
In the same way, we obtain $d= 3.9^{+0.7}_{-0.6}$ for Model 2.  
If we adopt a different set of chemical composition, the resultant distance is 
 also changed. For the $0.6~M_\odot$ WD we obtain 3.8 kpc for ($X,~Z)=$
(0.5, 0.01), 3.3 kpc for (0.7, 0.02) and 3.5 kpc for (0.7, 0.01). 
Taking into account that $Z=0.02$ is a bit larger than the recent estimate of 
solar value \citep[$Z=0.0128$:][]{gre08}, 3.3 kpc may be a smaller limit.
Thus, we may summarize our distance estimates as $d=3.8 \pm 0.7$ kpc. 
The error includes ambiguity of the $E(B-V)$, WD mass, and chemical composition.

\section{Discussion}
\label{sec_discussion}

As described in Section \ref{sec_model}, a nova becomes a supersoft 
X-ray source in the later phase of the outburst. 
In PU Vul, however, the maximum temperature is not high enough 
compared with a typical classical 
nova because of the less massive WD ($\sim 0.6 M_\odot$). 
Our theoretical model predicts that the 
outburst of PU Vul is still ongoing and the temperature is continuously rising. 
The temperature will finally reach the maximum temperature of 
$\sim 3 \times 10^5$ K just before hydrogen burning stops (X-ray turnoff), 
while the flux is almost constant at $2 - 3 \times 10^{37}$ erg~s$^{-1}$.   
Therefore, we may expect supersoft X-ray from PU Vul in future, but it  
is very difficult to predict this epoch, because it strongly depends 
on the model parameters, such as the optically-thin wind mass-loss rate, 
and the chemical composition of the envelope.

No X-ray observations of PU Vul have been made since November 1992. The 
detected X-ray flux was attributed to a shock origin of 
colliding winds \citep{hoa96,mue97}. PU Vul is a symbiotic nova in which 
the M-giant companion blows a massive cool wind that may preferentially 
distribute in the orbital plane. This cool wind may prevent a clear detection of 
supersoft X-rays from the hot WD. 
Therefore, the probability of detecting supersoft X-rays would 
be higher when the WD is in front of the cool wind of the giant (around 2014) 
and would be very low near the eclipse (2020-2021).

We have performed simulations for detectability of supersoft X-ray with  
the assumed values of $E(B-V)=0.43$ and $d=3.8$ kpc.
It will be possible to detect PU Vul at 0.007 cts~s$^{-1}$ 
with the EPIC-pn camera onboard {\it XMM-Newton} when the temperature 
is $\log T$ (K)$\sim 5.3$ and the luminosity is $\sim 5 \times 
10^{37}$ erg~s$^{-1}$ (probably in 2010--2020) and at about 1 cts~s$^{-1}$ 
near the X-ray turnoff time, i.e., $\log T$ (K)$\sim 5.5$ and 
$\sim 2 \times 10^{37}$ erg~s$^{-1}$ (probably in 2060--2090).
However, these estimates are based on our Model 1 and Model 2 only, 
and there is a large ambiguity due to uncertainty of the model parameters 
and assumptions. Especially the above expecting year depends strongly 
on the assumed wind-mass-loss rate after the photospheric 
temperature of the WD envelope rises to $\log T$ (K)$= 5.05$. 
If the wind-mass-loss will much weaken, the X-ray turnoff time 
is much later than the above estimates.

We expect that a large part of the envelope will remain on the WD 
after the outburst of PU Vul, because no optically thick wind occurs. 
From the initial envelope mass and the matter 
lost by the optically-thin wind as in Table \ref{table_model}, 
we estimate the envelope mass that will remain after the outburst.
Highly depending on the assumed mass-loss rate as well as 
the other parameters, it is estimated to be about 70--90 \% of the 
initial envelope mass for reliable 
model parameters. Mass-accreting WDs are, in general, potential candidates of 
progenitors of Type Ia supernovae (SNe Ia).  However, low mass WDs 
($\lesssim 0.9 M_\odot$) such as in PU Vul have not been considered as 
a candidate of SNe Ia, because it is hard to grow to Chandrasekhar mass limit 
\citep[e.g.][]{hkn99,hknu99,kat10}.

\section{Conclusions} \label{sec_conclusions}

Our main results are summarized as follows:

1. Based on the idea of \citet{kat09} that a long-lasted flat peak of 
 optical light curves can be reproduced by a sequence of 
wind-suppressed static-solutions, we have succeeded in reproducing the 
long-lasted optical flat peak of PU Vul
as well as the UV 1455 \AA~ continuum light curve. 
Our model is consistent with spectral features with no indication of 
strong winds in the flat peak of PU Vul.

2. An analysis of the {\it IUE} spectra of PU Vul indicates
  $E(B-V)= 0.43 \pm 0.05$.

3. We obtain a mass range of the WD between  0.55 and 0.65 $M_\odot$
by comparing our theoretical light curves with the UV light curve.

4. We obtain the distance of $d = 3.8 \pm 0.7$  kpc with $E(B-V)= 0.43 \pm 0.05$.

5.  We may conclude that the outburst of PU Vul is still on-going, and 
has already entered the supersoft X-ray phase. We encourage X-ray observations in 
$\sim 2014$, when the WD is in front of the red giant, that will provide important 
information on this symbiotic nova.

\acknowledgments 
We would like to thank Joanna Miko\l ajewska for valuable discussion  
on observational features of PU Vul. We are also grateful to the anonymous 
referee for useful comments that helped to improve the manuscript.
We also thank the American Association of Variable Star Observers (AAVSO) 
for the visual data of PU Vul.  
This research has been supported in part by the
Grant-in-Aid for Scientific Research (20540227,22540254) 
of the Japan Society for the Promotion of Science.

\end{document}